\documentclass[11pt]{article}

\def\be{\begin{equation}}
\def\ee{\end{equation}}
\def\bea{\begin{eqnarray}}
\def\eea{\end{eqnarray}}

\newcommand{\bdm}{\begin{displaymath}}
\newcommand{\edm}{\end{displaymath}}

\newcommand{\fs}{\hspace{0.5em}.}
\newcommand{\co}{\hspace{0.5em},}
\newcommand{\al}{&}
\newcommand{\eff}{{e\hspace{-0.1em}f\hspace{-0.18em}f}}

\newcommand{\QCD}{\mbox{\tiny Q\hspace{-0.05em}CD}}
\newcommand{\indR}{\mbox{\tiny R}}
\newcommand{\indL}{\mbox{\tiny L}}
\newcommand{\indV}{\mbox{\tiny V}}
\newcommand{\indP}{{\scriptscriptstyle P}}

\newcommand{\lvac}{\langle 0|\,}
\newcommand{\rvac}{\,|0\rangle}

\newcommand{\qbar}{\overline{\rule[0.42em]{0.4em}{0em}}\hspace{-0.45em}q}
\newcommand{\ubar}{\overline{\rule[0.42em]{0.4em}{0em}}\hspace{-0.5em}u}
\newcommand{\dbar}{\,\overline{\rule[0.65em]{0.4em}{0em}}\hspace{-0.6em}d}
\newcommand{\sbar}{\overline{\rule[0.42em]{0.4em}{0em}}\hspace{-0.5em}s}

\newcommand{\spc}{\hspace{2em}}
\newcommand{\bce}{\begin{center}}
\newcommand{\ece}{\end{center}}

\newcommand{\sss}{\scriptscriptstyle}
\newcommand{\za}{Z_{\!\sss{A}}}
\newcommand{\zm}{Z_{\!\sss{M}}}
\newcommand{\ga}{\gamma_{\!\sss{A}}}

\newcommand{\indtheta}{\mbox{\raisebox{-0.15em}{$\scriptstyle \theta$}}}

\begin{document}

\title{\bf Pseudoscalar decay constants at large {\boldmath $N_c$}}

\author{R. Kaiser, H. Leutwyler\\Institute for Theoretical Physics, University
  of Bern,\\  
Sidlerstr. 5, CH--3012 Bern, Switzerland\\
E-mail: kaiser@itp.unibe.ch, leutwyler@itp.unibe.ch}
\date{June 1998}
\maketitle
\begin{abstract}
  In the large $N_c$ limit, the variables required to analyze the
  low energy structure of QCD in the framework of an effective field theory
  necessarily include the degrees of freedom of the $\eta'$. We evaluate the
  decay constants of the pseudoscalar nonet to one loop within this extended
  framework and show that, as a consequence of the anomalous dimension of the
  singlet axial current, some of the effective coupling constants depend on
  the running scale of QCD. The calculation relies on a simultaneous expansion
  in powers of momenta, quark masses and $1/N_c$.
\end{abstract}

\section{Introduction}
The low energy properties of QCD are governed by an approximate,
spontaneously broken symmetry, which originates in the fact that three of the
quarks happen to be light. If $m_u,m_d,m_s$ are turned off, the symmetry
becomes exact. The spectrum of the theory then contains eight strictly massless
pseudoscalar mesons, the Goldstone bosons connected with the spontaneous
symmetry breakdown.

If the number of colours is taken large, the quark
loop graph which gives rise to the U(1)-anomaly 
is suppressed \cite{Large Nc}. This implies that, in the limit
$N_c\rightarrow\infty$, the singlet axial current is also conserved: The 
theory in effect acquires a higher degree of symmetry. Since the operator
$\qbar q$ fails to be invariant under the extra U(1)-symmetry, the formation
of a quark
condensate, $\lvac \qbar q\rvac\neq 0$, implies that this symmetry is also
spontaneously broken \cite{Coleman Witten}. The spectrum of QCD, therefore, 
contains a ninth
state, the $\eta^\prime$, which
becomes massless if not only $m_u,m_d,m_s$ are turned off, but if in addition
the number of colours is sent to infinity \cite{Minkowski}. 

Chiral symmetry imposes strong constraints on the
properties of the Goldstone bosons. These may be worked out in a
systematic manner by means of 
the effective Lagrangian method, which describes the low energy
structure of the theory in terms of an expansion in powers of
energies, momenta and quark masses \cite{Weinberg Physica,GL SU(3)}.
The fact that, in
the large $N_c$ limit, the $\eta'$ also plays the role of a Goldstone boson 
implies that the properties of this particle are subject to analogous
constraints, which may again be worked out by means of a suitable
effective Lagrangian. 

The main features of this
Lagrangian were studied long ago \cite{Leff U(3)}. In particular, it was shown
that the mass of the $\eta'$ is controlled by the topological susceptibility
of Gluodynamics and by the pion decay constant \footnote{Notation: We use the
 normalization 
$F_\pi=92.4$ MeV and absorb the coupling constant $g$ in the gluon field,
$D_\mu=\partial_\mu-iG_\mu$.}:
\be M_{\eta'}^2=6\,\frac{\tau_0}{F_\pi^2}+\ldots\ee
The topological susceptibility $\tau_0$ represents the mean square
winding number per unit volume of euclidean 
space, 
\bea 
\tau_0 \al\equiv\al\frac{\langle
\nu^{\,2}\rangle}{V}\hspace{-0.05em}
\mbox{\raisebox{0.4em}{\tiny GD}} 
=\int\!\! d^4x\lvac T\, \omega(x) \omega(0)\rvac_{\mbox{\tiny GD}}\co \\
\omega\al=\al\frac{1}{16\pi^2}\,\mbox{tr}
\hspace{-0.5em}\rule[-0.5em]{0em}{0em}_c\hspace{0.4em}
G_{\mu\nu}\tilde{G}^{\mu\nu}\co\hspace{3em}\nu=\int\!\! d^4x\,\omega
\fs\nonumber\eea
In the large $N_c$ limit, $\tau_0$ becomes independent of $N_c$ while $F_\pi$
grows in proportion to $\sqrt{N_c}$, so that $M_{\eta'}$ tends to 
zero \cite{Minkowski}. 

In more recent work, the derivative expansion of the relevant effective
Lagrangian was investigated beyond leading 
order \cite{Schechter Subbaraman Weigel}-\cite{Herrera
  2}. In particular, we have shown that the analysis of the
pseudoscalar decay constants in the framework of the effective theory 
requires two different mixing angles that are related through a low energy 
theorem \cite{Kaiser,Montpellier}. The decay 
constants are defined by
   \bea
   \lvac A_\mu^a | P \rangle = i p_\mu F_\indP^a \co \spc A_\mu^a=\qbar
   \gamma_\mu\gamma_5\mbox{$\frac{1}{2}$}\lambda^a q \co \label{fdef}
   \eea
where $ P\! =\!( \pi^0, \ldots, \eta') $ labels the pseudoscalar mesons. The
index $ a \!= \!(0, \ldots , 8)$ 
refers to a basis of U(3), normalized with
$\mbox{tr}(\lambda^a\lambda^b)=2\,\delta^{ab}$. The two angles mentioned
above specify the $\eta$ and $\eta'$
projections of the states $A_\mu^8\rvac$ and $A_\mu^0\rvac$, respectively:
\bea  F^8_{\;\eta}\al=\al\cos\vartheta_8\, F_8\co\hspace{1.8em}F^8_{\;\eta'}=
\sin \vartheta_8\,F_8\co\label{Feta}\\
F^0_{\;\eta}\al=\al  -\sin\vartheta_0\, F_0\co\hspace{1em}F^0_{\;\eta'}=\cos
\vartheta_0\, F_0\fs\nonumber\eea
Chiral symmetry implies that,
at leading order in a simultaneous expansion in powers of quark masses and
$1/N_c$, the difference between the two angles is determined by $F_K$ and 
$F_\pi$ \cite{Kaiser,Montpellier}:  
\be \sin
(\vartheta_0-\vartheta_8)=
\frac{2\sqrt{2}\,(F_K^{\;2}-F_\pi^{\;2})}{4F_K^2-F_\pi^2}
+\ldots\label{difference}\ee

The $\eta$-$\eta'$ mixing pattern has recently attracted much attention in
connection with production and decay processes involving these
particles  
\cite{Ball Frere
Tytgat}-\cite{CLEO}. 
As pointed out in \cite{Feldmann Kroll,Feldmann 
Kroll Stech}, the analysis in terms of two different mixing angles indeed
yields a more coherent picture than the
canonical treatment with $\vartheta_8=\vartheta_0$. The 
sign of the prediction is confirmed, 
but the numerical value obtained for the difference between the two
angles is somewhat smaller than what is required by (\ref{difference}). 
The purpose of the present paper is to discuss the
corrections which this low energy theorem 
receives from higher order effects. 

\section{Effective theory}
The dynamical variables of the QCD Lagrangian are the quark and gluon fields.
The low energy analysis of this system is based on an effective field theory
where the dynamical variables are mesonic fields 
with the quantum numbers of the Goldstone bosons.
For $N_c=3$, there are eight Goldstone fields -- the $\eta'$
occurs 
among the massive states which only show up indirectly, through their
contributions to the effective coupling constants. The octet of pseudoscalars 
may be collected in a unitary $3\times 3$--matrix $U(x)\in \mbox{SU(3)}$. 

The standard framework does not cover the large $N_c$ limit, 
however. Its domain of validity is
restricted by the condition \cite{Montpellier} 
\be m_s\,|\lvac\hspace{0.05em}\ubar\hspace{0.08em} u\rvac|< 9\, \tau_0\co\ee 
which is violated 
if the limit $N_c\rightarrow \infty$ is taken at fixed quark 
masses. Indeed, some of the standard ChPT formulae 
become meaningless in that limit \cite{Peris de  Rafael}.  
For the effective field theory to properly describe the low energy structure 
also when $N_c$ is taken large, the set of dynamical variables needs to be 
enlarged by adding a field that describes the extra Goldstone boson, the
$\eta'$. Quite generally, the effective fields live on the coset space G/H, 
where G is the
symmetry group of the Hamiltonian and H is the subgroup that leaves the vacuum
invariant. At large $N_c$, we have
$\mbox{G}=\mbox{U(3)}_{\indR}\times\mbox{U(3)}_{\indL}$,
$\mbox{H}=\mbox{U(3)}_{\indV}$, so that $U(x)\in\mbox{U(3)}$.
The extension from SU(3) to U(3) shows up in the phase of the determinant
\be \det U(x)=e^{i\psi(x)}\fs\ee
The field $\psi(x)$ describes the $\eta'$. 

The effective Lagrangian contains the meson 
field $U(x)$ and its
derivatives, ${\cal L}_{\eff}={\cal L}_{\eff}(U,\partial U, \partial^2
U,\ldots)$. In order to analyze the matrix elements of the quark currents, 
it is convenient to ab initio introduce corresponding external fields, 
replacing the QCD Lagrangian by
\be {\cal L}_{\QCD}= 
{\cal L}^0_{\QCD} +\qbar\gamma^\mu(v_\mu+\gamma_5 a_\mu) q-
\qbar(s-i\gamma_5 p)q -\theta\, \omega\fs\ee
The term ${\cal L}^0_{\QCD}$ describes the limit where the
masses of the three light quarks and the vacuum angle are set to 
zero. The external fields $v_\mu(x)$, $a_\mu(x)$, $s(x)$, $p(x)$ represent
hermitean $3\times 3$
matrices in flavour space. The mass matrix of the three light quarks
is contained in the scalar external field
$s(x)$. The vacuum angle $\theta(x)$ represents the variable conjugate
to the winding number density $\omega(x)$.

Naturally, ${\cal L}_{\eff}$ then also depends on the external fields 
and their derivatives. The dependence is strongly constrained by the
symmetries of the underlying theory \cite{GL SU(3),found}: 
Apart from the Wess-Zumino-Witten term, 
the effective Lagrangian is
invariant under chiral gauge transformations,
\bea \label{trafo}
\al\al r_\mu'= V_{\indR} r_\mu V_{\indR}^\dagger + i V_{\indR} \partial_\mu
V_{\indR}^\dagger\co\hspace{2em} 
l_\mu'= V_{\indL} l_\mu V_{\indL}^\dagger + i V_{\indL} \partial_\mu
V_{\indL}^\dagger\nonumber\\
\al\al s'+ip'= V_{\indR}(s+ip)V_{\indL}^\dagger\co\hspace{2.4em} U'=
V_{\indR} U V_{\indL}^\dagger\\
\al\al \theta'= \theta +i\ln\det V_{\indR}-i\ln\det V_{\indL} \co\nonumber\eea
with $r_\mu=v_\mu+ a_\mu$, $l_\mu=v_\mu-a_\mu$ and
$V_{\indR}(x), 
V_{\indL}(x)\in\mbox{U(3)}$.

The expansion of the general effective 
Lagrangian in powers of derivatives and quark masses starts 
with \cite{GL SU(3)} 
\bea
{\cal L}_{\eff}\al=\al-V_0+
V_1\langle D_\mu U^\dagger D^\mu U\rangle +
V_2 \langle (s+ip)\, U^\dagger \rangle + 
V_2^\star\langle (s-ip)\, U \rangle\nonumber
\\\al\al+ V_3 D_\mu\psi D^\mu\psi + 
V_4 D_\mu \psi D^\mu\theta +V_5 D_\mu\theta D^\mu\theta+O(p^4)\co\eea
where $\langle \ldots \rangle$ stands for the trace. The covariant derivatives 
are defined by
\bea D_\mu U\al=\al \partial_\mu U-i(v_\mu+a_\mu)\, U +i U\,(v_\mu-a_\mu)
\nonumber\\
 D_\mu\psi\al=\al\partial_\mu \psi -2\langle a_\mu\rangle\\
 D_\mu\theta\al=\al\partial_\mu \theta +2\langle a_\mu\rangle\fs\nonumber\eea
The transformation law (\ref{trafo}) implies that the
combination 
\be \tilde{\psi}\equiv\psi + \theta \ee
of the singlet field $\psi$ and the vacuum angle $\theta$ is invariant under 
chiral transformations. Chiral symmetry does therefore not constrain the
dependence of the Lagrangian on this variable: At this stage, the coefficients
are arbitrary functions thereof, $V_n=V_n(\tilde{\psi})$. They may be viewed
as potentials that control the dynamics of the singlet field $\psi(x)$.

As pointed out in ref. \cite{GL SU(3)}, the form of the potentials depends on
the choice of field variables: A transformation of the type $U\rightarrow
U\exp i f(\tilde{\psi})$ leaves the structure of the Lagrangian invariant, but
modifies the potentials. For our purposes it is convenient to exploit
this freedom by eliminating the term \footnote{The advantage of the
convention $V_4=0$ is that the quadratic part of the
Lagrangian then takes the simple form
${\cal L}_{\eff}=\frac{1}{2}A^{ab}(\partial_\mu\phi_a-a_{\mu a})
(\partial^\mu\phi_b-{a^\mu}_b
)-
\frac{1}{2}B^{ab}\phi_a\phi_b+\ldots$}
\be V_4=0\fs\ee
Moreover, we may discard $V_5$, because this term does not contribute to the
meson masses, decay constants or photonic transition rates.

In the 
limit $N_c\rightarrow\infty$, the potentials are dominated by the leading terms
of their Taylor series in powers of $\tilde{\psi}$ 
(for a discussion of the large $N_c$
counting rules, see ref. \cite{GL SU(3),Herrera 1}). 
Discarding terms that disappear in the large $N_c$ limit, the potentials
reduce to a set of constants: \footnote{At leading order of the
$1/N_c$--expansion, $\tau$ coincides with the
topological susceptibility of Gluodynamics: $\tau=\tau_0+O(1/N_c)$.} 
\bea
V_0\al=\al\mbox{const.}+\mbox{$\frac{1}{2}$}\,\tau\,
\tilde{\psi}^2+O(N_c^{-2})\co\nonumber\\
V_1\al=\al\mbox{$\frac{1}{4}$}F^2+O(N_c^{-1})\co\\
V_2\al=\al\mbox{$\frac{1}{2}$}F^2B\left(1-
\mbox{$\frac{1}{3}$}\,i \Lambda_2\,\tilde{\psi}\right)
+O(N_c^{-1})\co\nonumber\\
V_3\al=\al\mbox{$\frac{1}{12}$}F^2 \Lambda_1+O(N_c^{-2})\nonumber\fs\eea
At this order, the effective Lagrangian thus contains five coupling constants:
\be F^2= O(N_c)\,;\hspace{2em}B,\,\tau=O(1)\,;\hspace{2em}\Lambda_1,\,
\Lambda_2=O(N_c^{-1})\fs
\ee
The constant $B$ only occurs together with the scalar and pseudoscalar
external fields, through the combination
\be \chi\equiv 2B(s+ip)\fs\ee

We add a remark concerning the comparison between the extended 
effective theory based on $U(x)\in\mbox{U(3)}$ and the standard framework,
where $U(x)\in\mbox{SU(3)}$. As discussed in detail in
ref. \cite{GL SU(3)}, the Ward identities obeyed by the Green functions of the
vector and axial currents imply that the SU(3) effective Lagrangian is 
invariant under the full group U(3)$_{\indR}\times$U(3)$_{\indL}$ of chiral 
gauge transformations -- up to the Wess-Zumino-Witten term, which accounts for
the anomalies occurring in these identities. 
The transformation law
$U'=V_{\indR}UV_{\indL}^\dagger$, however, violates the
constraint $\det U=1$. Instead of modifying the transformation law, it is more
convenient to perform a change of variables, 
$U\rightarrow U\exp (-\frac{i}{3}\theta)$, i.e. to modify the constraint, which
then takes the form
$\det U=e^{-i\theta}$. The condition amounts to the gauge invariant 
relation $\psi(x)=-\theta(x)$, 
which states that the standard framework arises from the extended one if the 
extra variable $\psi(x)$ is fixed at the minimum of the potential
$V_0=V_0(\psi+\theta)$.  

\section{Higher orders in the derivative expansion}
At order $p^4$, the general effective Lagrangian contains altogether 57
potentials\footnote{ 
Some of these may be eliminated with a suitable change of 
variables. In
particular, the terms $O_{46}$ and $O_{52}$ -- which do
generate a contribution 
to the decay constants -- are removed with a transformation of the
form $U\rightarrow U\,\exp[f_1\, D_\mu \theta\, U^\dagger D^\mu U  
 +i\,f_2\, D_\mu D^\mu \theta]$. 
We exploit this freedom and set $L_{46}=L_{52}=0$.}, 
which are listed explicitly in ref. \cite{Herrera 1}. Expanding
these in powers of $\tilde{\psi}$ and retaining only
those terms that (a) generate a contribution to the masses or decay 
constants and (b) do not disappear in the 
limit $N_c\rightarrow\infty$, the
general expression boils down to
\begin{eqnarray}\label{L4}
{\cal L}^{(p^4)}\al=\al L_4 \langle D_\mu U^\dagger D^\mu U\rangle
\langle \chi^\dagger U + U^\dagger \chi\rangle+
L_5 \langle D_\mu U^\dagger D^\mu U (
\chi^\dagger U + U^\dagger \chi)\rangle\nonumber\\\al\al+
L_6\langle \chi^\dagger U + U^\dagger \chi\rangle^2 +
L_7\langle \chi^\dagger U - U^\dagger \chi\rangle^2\\\al\al +
L_8\langle\chi^\dagger U\chi^\dagger U+U^\dagger\chi U^\dagger \chi\rangle
+L_{18}\,iD_\mu \psi\,\langle D^\mu U^\dagger \chi-D^\mu U\chi^\dagger\rangle
\nonumber\\\al\al
+L_{25}\,i\,\tilde{\psi}\,\langle U^\dagger\chi U^\dagger\chi-\chi^\dagger
U\chi^\dagger U\rangle+O(N_c^{-1})\fs\nonumber\eea
The coupling constants $L_4,\,\ldots\,,\,L_8$ are
familiar from the standard framework, where the effective field $U(x)$ is an
element of the group SU(3) rather than U(3). Indeed, if the singlet component
of the meson field is
integrated out, the above expression for the effective Lagrangian does take
the standard form. Note, however, that the effective coupling 
constant $L_7$ then picks up an
extra contribution from $\eta'$-exchange, which in the 
large $N_c$
limit even dominates: 
\be\label{L7} L_7^{\mbox{\tiny
    SU(3)}}=-\frac{F^4(1+\Lambda_2)^2}{288\tau}+L_7\fs\ee 
In our context, the extension from SU(3) to U(3) thus gives rise to two 
additional $p^4$-couplings, for which we have 
retained the numbering
used by Herrera-Sikl\'{o}dy et al. (in their notation: 
$L_{18}\equiv L_{18}(0)$, 
$L_{25}\equiv L_{25}'(0)$). The large $N_c$ counting rules 
imply \cite{Herrera 1} 
\be L_5,\,L_8=O(N_c)\,;\hspace{2em}L_4,\,L_6,\,L_7,\,L_{18},\,L_{25}
=O(1)\fs\ee

It is convenient to order the triple expansion in (i) the number of
derivatives, (ii) powers of quark 
masses and (iii) powers of $1/N_c$ by treating the three expansion parameters 
as small 
quantities of order
\be \partial_\mu
=O(\sqrt{\delta}\;)\co\hspace{2em}m=O(\delta)\co\hspace{2em}1/N_c=O(\delta)\fs
 \ee
In this bookkeeping, the fields $U$, $\psi$, $\theta$ are of order
$\delta^0$, while $v_\mu$, $a_\mu$ count as terms of $O(\sqrt{\delta})$ 
and $s$, $p$ are of $O(\delta)$. The expansion
of the effective Lagrangian
then takes the form 
\be {\cal L}_{\eff}=  
{\cal  L}^{(0)}+{\cal L}^{(1)} +{\cal L}^{(2)}+  
 \ldots\ee
It starts with the contributions of order $\delta^0$:
\be\label{L0}
{\cal L}^{(0)}=\mbox{$\frac{1}{4}$}F^2\langle D_\mu U^\dagger D^\mu U\rangle +
\mbox{$\frac{1}{4}$} F^2\langle \chi^\dagger U + U^\dagger \chi\rangle -
\mbox{$\frac{1}{2}$}\,\tau (\psi + \theta)^2\fs\ee
The term ${\cal L}^{(1)}=O(\delta)$ contains the contributions of $O(N_c p^4)$ 
as well as those of $O(p^2)$
generated by the Okubo-Iizuka-Zweig rule violating couplings 
$\Lambda_1$, $\Lambda_2$:
\bea {\cal L}^{(1)}\al=\al L_5 \langle D_\mu U^\dagger D^\mu U (
\chi^\dagger U + U^\dagger \chi)\rangle +
L_8\langle\chi^\dagger U\chi^\dagger U+U^\dagger\chi U^\dagger \chi\rangle
\\\al\al+
\mbox{$\frac{1}{12}$}F^2\Lambda_1D_\mu \psi D^\mu\psi
+\mbox{$\frac{1}{12}$}F^2 \Lambda_2 \,i(\psi+\theta)
\langle\chi^\dagger U-U^\dagger\chi\rangle
\fs\nonumber\eea
The remainder of the expression in eq. (\ref{L4}) 
belongs to ${\cal L}^{(2)}=O(\delta^2)$:
\bea
{\cal L}^{(2)}\al=\al L_4 \langle D_\mu U^\dagger D^\mu U\rangle
\langle \chi^\dagger U + U^\dagger \chi\rangle+
L_6\langle \chi^\dagger U + U^\dagger \chi\rangle^2
+L_7\langle \chi^\dagger U - U^\dagger \chi\rangle^2\nonumber
\\\al\al
+L_{18}\,iD_\mu \psi\,\langle D^\mu U^\dagger \chi-D^\mu U\chi^\dagger\rangle
+L_{25}\,i(\psi+\theta)\,\langle U^\dagger\chi U^\dagger\chi-\chi^\dagger
U\chi^\dagger U\rangle\nonumber\\\al\al +O(N_c\,p^6)\fs\eea

\section{Anomalous dimensions}
The effective theory is constructed in such a way that it reproduces the
derivative expansion of the effective action of QCD.
The effective action represents the connected 
vacuum-to-vacuum amplitude in the
presence of the external fields,
\be e^{iS_\eff}=\langle 0\;\mbox{out}|\,0\; \mbox{in}
\rangle_{v_\mu,a_\mu,s,p,\theta}\fs\ee
It collects the set of all connected Green functions formed with the operators
$\qbar \gamma_\mu q$, $\qbar \gamma_\mu\gamma_5 q$, $\qbar q$, $\qbar \gamma_5
q$, $\omega$ and depends on the running coupling constant $g$ as well as on the
renormalization scale $\mu$:
$S_{\eff}=S_{\eff}(v_\mu,a_\mu,s,p,\theta,g,\mu)$.  

In principle, all of the
effective coupling constants occurring in ${\cal L}_{\eff}$ are
determined by $g$, $\mu$ and by the masses of the heavy quarks $c$, $b$, $t$.
We now discuss the manner in which the effective coupling constants depend on 
the running scale of QCD. In particular, we wish to work out the consequences
of the fact that the matrix elements of the singlet axial current depend on the
renormalization, because this operator 
carries anomalous dimension. To our knowledge, this issue is not discussed
in the literature, because the matrix elements of the singlet axial current
are usually considered only at leading order of the $1/N_c$--expansion 
\footnote{Note that we are considering the scale dependence within QCD -- the
   one arising
from the logarithmic divergences that occur in the effective theory is an 
entirely
different issue (see section \ref{loops}).}.
 
For the Green functions to remain the same when the renormalization scale is
changed, the coupling constant must be adapted,
\be \mu \frac{dg}{d\mu}=-\beta(g)=-\beta_0\frac{g^3}{16\pi^2}+O(g^5)\co
\hspace{2em}\beta_0=\mbox{$\frac{1}{3}$}(11N_c-2N_f)\fs\ee 
This does not suffice, however, because some
of the operators under consideration carry anomalous dimension. In particular,
the scalar and pseudoscalar operators require renormalization
for their Green functions to remain unaffected by a change of
scale. The same holds for the Green functions containing the singlet axial
current $A^0_\mu$ -- this operator also receives multiplicative
renormalization \cite{Kodaira}. As it is the case with the quark masses, 
the values the decay constants associated 
with the singlet current ($F^0_\eta$ and $F^0_{\eta'}$) therefore 
depend on the scale. 

The external fields may be viewed as space-time dependent coupling constants.
It is convenient to compensate the scale dependence of the operators by
treating the corresponding external fields as scale dependent quantities, in
such a way that the effective action becomes scale independent. This requires
a renormalization of $s(x)$, $p(x)$ and of the singlet component of the axial
external field. On the other hand, the algebra
obeyed by the charges of ${\rm SU(3)} \times {\rm SU(3)}$ implies that the
octet components of the vector and axial currents are of canonical
dimension, so that their matrix elements are scale independent. The same is
true of the singlet vector current (baryon number). To simplify the
renormalization group behaviour of the external fields,
it is convenient to replace $ s $ and $ p $ by the combination 
\be   m_{\indtheta} = e^{\frac{i}{3}\theta}(s+ip) \co\ee 
because this quantity is invariant under the
transformations generated by the singlet charges. 

The various Green functions are Lorentz invariant expressions formed with the
external momenta. There are two categories: those that may be
expressed in terms of the metric $g_{\mu\nu}$ alone (natural parity) and those
for which the expression is linear in the tensor $\epsilon_{\mu\nu\rho\sigma}$
(unnatural parity). Accordingly, the effective action may be decomposed into
two parts that are of natural and unnatural parity, respectively. The
anomalies only show up in the latter. As we are restricting ourselves to an
analysis of the decay constants, we are concerned with the natural parity part,
which is invariant under the transformation (\ref{trafo}).

The singlet component of $a_\mu$ is given by the trace $\langle a_\mu\rangle$.
Denoting the octet part by $\hat{a}_\mu\equiv 
a_\mu-\frac{1}{3}\langle a_\mu\rangle$
and using the variable $m_{\indtheta}$ instead of $s$, 
$p$, the arguments of the effective action are: 
$v_\mu,\hat{a}_\mu,\langle a_\mu\rangle,m_{\indtheta},\theta$. 
Invariance under the transformation
(\ref{trafo}) implies
that $\langle a_\mu\rangle$ and $\theta$ only enter through the gauge
invariant combination \footnote{The statement 
represents a generalization of the familiar fact
that the quark mass matrix ${\cal M}=\mbox{diag}(m_u,m_d,m_s)$ and the
vacuum angle only 
enter in the combination $e^{\frac{i}{3}\theta}\,{\cal M}$.}
$D_\mu \theta=\partial_\mu \theta + 2\langle a_\mu\rangle$, so that
$S_{\eff}=S_{\eff}(v_\mu,\hat{a}_\mu,D_\mu\theta,m_{\indtheta},g,\mu)$. 
The symmetries of
the theory thus  
imply that $\partial_\mu \theta$ picks up the same renormalization factor
as the singlet axial current: The effective action is invariant under a
change of the running scale, provided the external fields are
subject to the transformation
   \begin{eqnarray}
   m_{\indtheta} \to  \zm^{-1}\,m_{\indtheta} \co \spc
   D_\mu\theta  \to \za^{-1} D_\mu \theta \fs
   \end{eqnarray}
$ \zm $ is the familiar factor that describes the scale dependence of
the quark masses. The term $ \za $ specifies the one of the singlet decay
constants:  
\be F^0_\indP \to \za F^0_\indP \fs \ee 
It is determined by the anomalous dimension $\ga $ of the singlet axial
current \cite{Kodaira} 
   \begin{eqnarray}
   \mu \frac{d \za }{d \mu} = \ga \za \co \spc
   \ga  = - \frac{3 N_f (N_c^2-1)}{8\,N_c}\left(\frac{g}{2
   \pi}\right)^4 + 
   O(g^6) \fs \label{anodim}  
   \end{eqnarray}
While the scale dependence of the
coupling constant and of the quark masses shows up already at leading order in
the $1/N_c$--expansion, the triangle graph responsible for the anomalous
dimension of the singlet axial current is suppressed by one power of $1/N_c$,
so that $\za=1+O(1/N_c)$.

\section{Scaling laws for the effective coupling constants} 
We now translate these properties into corresponding scaling laws for the
coupling constants of the effective Lagrangian. This may be done by working
out the first few terms in the chiral expansion of suitable observables.
The leading term, $F$, is evidently scale independent: It represents the pion
decay constant in the 
chiral limit. On the other hand, the lowest order result for the pion mass,
$M_\pi^2=B(m_u+m_d)+O(m^2)$, shows that $B$ transforms contragrediently to the
quark masses:
\be B\rightarrow \zm B\fs\ee
Next, consider the chiral limit and set $v_\mu=\hat{a}_\mu=\theta=0$, where
the classical solution (tree graphs) takes the form $U=e^{\frac{i}{3}\psi}{\bf
  1}$. The terms linear and quadratic in $\psi$,
\be \label{Lpsi}
{\cal L}_{\eff}=\mbox{$\frac{1}{12}$}F^2(1+\Lambda_1) D_\mu\psi D^\mu\psi
-\mbox{$\frac{1}{2}$}\tau\, \psi^2+\ldots \ee
yield the singlet 
decay constant and the mass of the $\eta'$ at tree level
\be  F^0_{\eta'}= \sqrt{1+\Lambda_1}
\,F\co\hspace{2em} M_{\eta'}^2=
\frac{6\tau}{(1+\Lambda_1)\, F^2}\fs\nonumber\ee
Hence $\tau$ and $\Lambda_1$ scale with \footnote{The scale-dependence of
  $\tau$ only shows up at nonleading order -- since the triangle
  graph responsible for $\za$ does not occur in Gluodynamics, the topological
  susceptibility $\tau_0$ is independent of the running scale.}
\be \tau\rightarrow \za^2\tau\co\hspace{2em}1+\Lambda_1\rightarrow
\za^2(1+\Lambda_1) \fs\ee
The coupling constants $L_4,\ldots\,,L_8$ are related to Taylor
coefficients occurring in the expansion of $F_\pi$, $F_K$, $M_\pi^2$, $M_K^2$ 
in powers of the quark masses and are thus independent of the QCD-scale.
Finally, the evaluation of the decay constants and masses of $\eta$ and $\eta'$
shows that the remaining coupling constants scale according to
\bea 1+\Lambda_2\al\rightarrow\al \za(1+\Lambda_2)\co\hspace{2em}\nonumber\\
2L_5+3L_{18}\al\rightarrow\al \za (2L_5+3L_{18})\co\hspace{2em}\\
2L_8-3L_{25}\al\rightarrow\al \za (2L_8-3L_{25})\fs\nonumber\eea
As a check, we note that the scaling laws for $\tau$ and $\Lambda_2$ 
are consistent with the 
scale-independence of the coupling constant $L_7^{\mbox{\tiny SU(3)}}$
specified in (\ref{L7}).

The net result of this analysis is the following. Let us replace the
external fields $s$ and $p$ by the variable $m_{\indtheta}$, also in the
effective Lagrangian and use the scale independent quantity
\be \chi_{\indtheta}= e^{\frac{i}{3}\theta}\chi\ee
instead of $\chi=2B(s+ip)$. Furthermore, to explicitly 
display the dependence on the
singlet component of the effective field, we set 
$U=e^{\frac{i}{3}\psi}\hat{U}$, 
so that $\hat{U}\in \mbox{SU(3)}$. The effective Lagrangian 
then depends on the fields $\hat{U},\psi+\theta,v_\mu,\hat{a}_\mu,
 D_\mu\theta,\chi_{\indtheta}$ and their derivatives. \footnote{In particular,
   we have $D_\mu\psi=\partial_\mu(\psi+\theta)-D_\mu\theta$.}
The vacuum angle only enters together with $\psi$ --
this is what becomes of $\theta$-independence at the level of the effective
theory. The above scaling laws amount to the statement that ${\cal L}_{\eff}$
is invariant under a change of the QCD scale, provided the singlet component
of the effective field is renormalized according to
\be \psi+\theta\rightarrow \za^{-1}\,(\psi+\theta)\fs\ee
Conversely, this property ensures that the effective action generated by
the effective theory is scale independent.

\section{Loops and chiral logarithms}\label{loops}
If the low energy expansion is ordered according to powers of $\delta$,
graphs containing $\ell$ loops yield contributions that are at most 
of order 
$\delta^{2\ell}$. Up to and including $O(\delta^2)$, only the one-loop 
graphs generated by the leading
term ${\cal L}^{(0)}$ in eq. (\ref{L0}) contribute. 

The divergences occurring in the one-loop graphs are absorbed in a
renormalization of the effective coupling constants. The relevant
coefficients have been worked out in ref. \cite{Herrera 1}. The main difference
to the standard framework is that loops involving the propagation of an 
$\eta'$ require a renormalization of the constant $B$: In dimensional
regularization this constant contains a pole at $d=4$. Expressed in terms of 
the factor
\be
\lambda =\frac{\mu_\chi^{d-4}}{16\pi^2}\left\{\frac{1}{d-4}-\frac{1}{2}
\left(\rule[1em]{0em}{0em}\ln 4\pi+\Gamma'(1)+1\right)\right\}\ee
the renormalization reads (to distinguish it from the running scale $\mu$ 
of QCD, we denote the scale used in the renormalization of the effective
theory by $\mu_\chi$) 
\be B= B^r\!\!\left\{1+\frac{4\tau}{F^4}\lambda 
\right\}\fs\ee
Since $\lambda$ depends on the renormalization scale $\mu_\chi$, 
the same is true of the renormalized coupling
constant (note that the running scale $\mu$ of QCD is kept fixed):
\be\label{B} \mu_\chi\frac{\partial B^r}{\partial\mu_\chi}=-\frac{\tau B^r}{4\pi^2F^4}\fs\ee
 The remaining couplings
occurring
at order $p^0$ and $p^2$ ($F$, $\tau$,
$\Lambda_1$ and $\Lambda_2$) do not pick up renormalization, while the one for
the terms of order $p^4$ is of the standard form, but the coefficients
$\Gamma_6$ and $\Gamma_8$ differ
from those relevant for the SU(3)-framework \cite{Herrera 1}
\bea\label{scaling} L_n\al=\al L_n^r+\Gamma_n\lambda\co\\
\Gamma_4\al=\al\mbox{$\frac{1}{8}$},\;\;\Gamma_5=\mbox{$\frac{3}{8}$},\;\;
\Gamma_6=\mbox{$\frac{1}{16}$},\;\;\Gamma_7=0,\;\;\Gamma_8=
\mbox{$\frac{3}{16}$},\;\;
\Gamma_{18}= -\mbox{$\frac{1}{4}$},\;\;\Gamma_{25}=0\fs\nonumber\eea

The evaluation of the loops yields the following expressions for the decay
constants (we disregard isospin breaking, setting $m_u=m_d$):
\bea F_\pi\al=\al F\!\left\{1+4(2M_K^2+M_\pi^2)F^{-2}L_4^r
+4 M_\pi^2 F^{-2}L_5^r-2\mu_\pi -\mu_K+O(p^4)\right\}\co\nonumber\\
 F_K\al=\al F\!\left\{1+4(2M_K^2+M_\pi^2)F^{-2}L_4^r+4 M_K^2
  F^{-2}L_5^r\right.\\
\al\al \hspace{7em}\left.-\mbox{$\frac{3}{4}$}
(\mu_\pi +2\mu_K+\cos^2\!\vartheta
  \mu_\eta + \sin^2\!\vartheta\mu_{\eta'})+O(p^4)\right\}\co
\nonumber\eea
where $\mu_P$ is the standard chiral logarithm,
\be \mu_P=\frac{M_P^2}{32\pi^2F^2}\ln\frac{M_P^2}{\mu_\chi^2}\fs\ee
One readily checks that the scale dependence of the 
chiral logarithms cancels the one of the effective coupling constants $L_4^r$,
$L_5^r$, provided the masses and the mixing angle 
$\vartheta$ are evaluated at leading order
of the expansion, where
\bea \label{mf}\cos^2\!\vartheta M_\eta^2+\sin^2\!\vartheta
M_{\eta'}^2=\frac{1}{3}\,(4M_K^2-M_\pi^2)\co \\
\sin 2\vartheta=-\frac{4\sqrt{2}}{3}\,
\frac{M_K^2-M_\pi^2}{M_{\eta'}^2-M_\eta^2}\fs\nonumber
\eea
These lowest order relations amount to a constraint on the pseudoscalar masses,
which the observed values do not obey \cite{Georgi}. For definiteness, we
evaluate the one loop formulae with the physical values of $M_\pi$, $M_K$ and
$M_{\eta'}$ (the above relations then yield $\vartheta\simeq -20^\circ$,
$M_\eta\simeq M_K$). 

\section{Results}
The result for $F_\pi$ is identical with the one found in
the framework of SU(3), where loops involving the propagation of an $\eta'$ do
not occur. In the case of $F_K$, however, there is a contribution from an
$\eta'$-loop. The relation 
\be\label{FKpi}\frac{F_K}{F_\pi}=1+4(M_K^2-M_\pi^2)F^{-2}L_5^r+
\mbox{$\frac{5}{4}$}
\mu_\pi -\mbox{$\frac{1}{2}$}\mu_K-\mbox{$\frac{3}{4}$}(\cos^2\!\vartheta
  \mu_\eta + \sin^2\!\vartheta\mu_{\eta'})\ee
differs from the corresponding SU(3) formula:
$\mu_\eta$ is replaced by the term
$(\cos^2\!\vartheta
  \mu_\eta + \sin^2\!\vartheta\mu_{\eta'})$. Numerically,
 the difference is not
significant, however: The two terms 
differ by less
  than $0.02$. 

The constants $F_8$ and $F_0$ are defined through the relations (\ref{Feta}):
\be(F_8)^2= (F_\eta^8)^2 + (F_{\eta'}^8)^2\co\hspace{2em} 
(F_0)^2= (F_\eta^0)^2 + (F_{\eta'}^0)^2\nonumber\fs\ee
For $F_8$, the evaluation of the loop graphs yields
\be\label{F8} F_8 =
F\left\{1+4(2M_K^2+M_\pi^2)F^{-2}L_4^r+
\mbox{$\frac{4}{3}$}(4M_K^2-M_\pi^2)F^{-2}L_5^r
  -3\mu_K +O(p^4)\right\}\fs\ee
Eliminating the coupling
constant $L_5^r$ with (\ref{FKpi}), we obtain a parameter free prediction for 
$F_8$:
\be \frac{F_8}{F_\pi}=1+\mbox{$\frac{4}{3}$}\,\Delta_F
+\mbox{$\frac{1}{3}$}\mu_\pi-\mbox{$\frac{4}{3}$}\mu_K+
\cos^2\!\vartheta
  \mu_\eta + \sin^2\!\vartheta\mu_{\eta'}\co\ee
with $\Delta_F\equiv F_K/F_\pi-1=0.22$. 
Numerically, this yields $F_8=1.34\, F_\pi$.

Since $F_0$ depends on the running scale
of QCD, chiral symmetry does not predict its numerical value.
In the chiral limit, we have 
$F_0=\sqrt{1+\Lambda_1}\,F$, in agreement with the scaling laws for $F_0$ and
$\Lambda_1$. The corrections of order $m_s$ are determined
by the coupling constants $L_4=O(1)$, $L_5=O(N_c)$ and $L_{18}=O(1)$.
In contrast to $F_8$, the constant $F_0$ does not receive contributions from
loops. The result of the tree graph calculation may be written in 
scale invariant form:
\bea\label{F0} F_0\al=\al
\sqrt{1+\Lambda_1}\,\bar{F}_0\nonumber\co\\
\bar{F}_0\al=\al F\left\{1
+\mbox{$\frac{4}{3}$}\,(2M_K^2+M_\pi^2)F^{-2}\,(3L_4^r-L_5^r+L_A) 
+O(p^4)\right\}\fs\label{F0pi}\eea
The term $L_A$ stands for the combination
\be L_A\equiv \frac{2L_5^r+3L_{18}^r}{\sqrt{1+\Lambda_1}}\co\ee
which is independent of the running scale $\mu$ of QCD as well as of the 
scale $\mu_\chi$ used in the renormalization of the effective theory 
($2\,\Gamma_5+3\,\Gamma_{18}=0$). In view of $3\,\Gamma_4-\,\Gamma_5=0$, 
this also holds for $\bar{F}_0$. 
In the ratio $\bar{F}_0/F_\pi$, the coupling constant $L_4^r$ drops out. Hence
$\bar{F}_0$ may be expressed in terms of the scale invariant quantities 
$F_\pi,F_K$ and $L_A$.  

In the large $N_c$ limit, $L_A$ is dominated by the contribution
from $L_5^r$ -- both $L_{18}^r$ and $\Lambda_1$ represent 
OZI-violating corrections. The coupling constant $L_A$ also determines the 
slope of the scalar form factor
$\langle\eta(p')|\,\ubar u +\dbar d-2 \sbar s\,|\eta(p)\rangle$,
but there is no experimental information about it and, to our knowledge,
 a dispersive calculation
has not been performed, either. The resonance exchange
calculations described in \cite{EGPdeR} yield $L_{18}=\Lambda_1=0$,
because the resonance couplings used there obey the OZI-rule. Inserting this
estimate at the scale $\mu=M_\rho$, we obtain  
$L_A\simeq 2L_5^r(M_\rho)$, so that $\bar{F}_0\simeq F_\pi$.
Note, however, that the result is very sensitive to the scale 
at which the
OZI-rule ($|L_{18}^r(\mu_\chi)|\ll |L_5^r(\mu_\chi)|$, $|\Lambda_1|\ll 1$) is 
assumed to be valid. The phenomenological 
determination of $\bar{F}_0$ on the basis of the photonic decays of $\eta$ 
and $\eta'$ should yield a more reliable value \cite{KL}. 

The above effective Lagrangian also allows us to calculate the two angles
$\vartheta_8$, $\vartheta_0$ as well as
$M_\pi,M_K,M_\eta,M_{\eta'}$ in terms of the coupling constants occurring
therein. We will report about this calculation elsewhere \cite{KL} and restrict
ourselves to the result for the difference between the two angles, which is
related to
\be F_\eta^8 F_\eta^0+F_{\eta'}^8
     F_{\eta'}^0=-F_8F_0\sin(\vartheta_0-\vartheta_8)\fs\ee
Since this combination of decay constants does not receive contributions from
loops, it suffices to work out the tree graphs of the
effective Lagrangian. The result may be written in the scale invariant form
\be
\sin(\vartheta_0-\vartheta_8)=
\frac{8\sqrt{2}(M_K^2-M_\pi^2)L_A}{\rule{0em}{1em}3 F_8 \bar{F}_0}
+O(p^4)\fs\label{d08}\ee  
The formula includes all corrections of order $1/N_c$, but accounts
for the symmetry breaking effects generated by the quark masses only to 
leading order. With the rough estimate for $L_A$ discussed above, the result
amounts to $\vartheta_0-\vartheta_8\simeq 14^\circ$, to be compared with the
number that follows from the leading order formula (\ref{difference}):
$\vartheta_0-\vartheta_8\simeq 16^\circ$. This may indicate that the higher 
order effects tend to slightly reduce the difference between the two angles,
but phenomenological input is required to arrive at a reliable result. 
In this context, the relations (\ref{F0pi}) and (\ref{d08}) are very useful,
because, even if the coupling constant $L_A$ is treated as a free parameter,
they correlate the 
difference between the two angles with the magnitude of $\bar{F}_0$.

\section*{Acknowledgements}
Part of this work was done while one of us (H.\ L.) stayed at the CSSM. He
wishes to thank the organizers of the workshop for hospitality and support. 
Moreover, we thank Hans Bijnens, Thorsten Feldmann and Bachir Moussallam for 
informative discussions and comments. 


\end{document}